# Breather arrest in a chain of damped oscillators with Hertzian contact

Matteo Strozzi and Oleg V. Gendelman*

Faculty of Mechanical Engineering, Technion - Israel Institute of Technology,
Technion City, Haifa 3200003, Israel
* Corresponding Author, email: ovgend@technion.ac.il

**Abstract**
We explore breather propagation in the damped oscillatory chain with essentially nonlinear (nonlinearizable) nearest-neighbour coupling. Combination of the damping and the substantially nonlinear coupling leads to rather unusual two-stage pattern of the breather propagation. The first stage occurs at finite fragment of the chain and is characterized by power-law decay of the breather amplitude. The second stage is characterized by extremely small breather amplitudes that decay hyper-exponentially with the site number. Thus, practically, one can speak about finite penetration depth of the breather. This phenomenon is referred to as breather arrest (BA). As particular example, we explore the chain with Hertzian contacts. Dependences of the breather penetration depth on the initial excitation and on the damping coefficient on the breather penetration depth obey power laws. The results are rationalized by considering beating responses in a system of two damped linear oscillators with strongly nonlinear (non-linearizable) coupling. Initial excitation of one of these oscillators leads to strictly finite number of beating cycles. Then, the beating cycle in this simplified system is associated with the passage of the discrete breather between the neighbouring sites in the chain. Somewhat surprisingly, this simplified model reliably predicts main quantitative features of the breather arrest in the chain, including the exponents in numerically observed power laws.

## 1. Introduction

Spatially localized oscillations propagating in nonlinear lattices, such as solitary waves and discrete breathers, have attracted a lot of interest in recent years. This interest has been driven both by purely theoretical questions, as well as by versatile possible applications including electro-mechanical devices, such as shock and energy absorbers [1-4], actuators and sensors [5-6], acoustic lenses and diodes [7-8].

In this paper, we address the dynamics of discrete breathers in particular type of damped nonlinear lattice. Discrete breathers are described as spatially localized time-periodic oscillations arising in nonlinear lattices due to the simultaneous presence of spatial discreteness, nonlinearity and local potentials [9-11]. Vibro-impact lattices also represent an example of nonlinear lattices bearing discrete breathers, with nonlinearity stemming from rigid impact constraints [12-15].

Many studies devoted to the discrete breathers consider Hamiltonian dynamics. However, in practical applications, dissipation cannot be neglected and therefore external/internal excitations should be applied to counter-balance the energy dissipation due to inelastic impacts and to preserve the discrete breather [9]. The energy dissipation transforms the conservative dynamics of the Hamiltonian lattice to the non-stationary dynamics of a dissipative damped-forced system [10]. Most non-conservative, multi-degree-of-freedom (multi-DOF) nonlinear vibro-impact systems possessing discrete breathersy can be analysed only by numeric means or by approximate analytical methods [9,16]. Conversely, there are very few damped-forced dissipative models of vibro-impact lattices where some exact analytical solutions for the discrete breathers can be derived [14,17-19].

If the external excitation and the local damping balance the effects of spatial discreteness and nonlinearity, one can observe the discrete breathers propagating along the lattice; these objects are referred to as travelling breathers [20-21]. Several analytical and numerical studies address the travelling breathers in different types of nonlinear lattices, e.g. Klein-Gordon lattices [22-24], Fermi-Pasta-Ulam lattices [25-27] and granular lattices [28-30].

The presence of local damping without external forcing inevitably leads to the energy dissipation and therefore decay of the amplitude of the propagating breather. Recently it was demonstrated [31] that if the coupling between the neighbours is essentially nonlinear (i.e. nonlinearizable), then, due to peculiar interaction between these two factors (essential nonlinearity and damping) one can observe an interesting phenomenon of *breather arrest* (BA). Based on preliminary results of [31], and, more significantly, on the findings presented below, the latter is defined as the abrupt switch from power-law to hyper-exponential decay of the maximum breather amplitude as a function of the site in the chain. After the crossover, the breather amplitude becomes extremely small, and one can speak about penetration to *finite* depth in the lattice. This result has been obtained in [31] for somewhat exotic (albeit physically realizable in principle) case of purely cubic coupling.

Main goal of current study is generalization and rationalization of these results, and, in particular, the exploration of the BA phenomenon in a chain with much more realistic and ubiquitous Hertzian coupling. This coupling is strongly nonlinear and purely repulsive, and vanishes in absence of the contact, i.e. under extension. Hertzian contact forces attract a lot of attention, since they allow the modelling and exploration of nonlinear dynamical processes in the systems with emerging contacts and unilateral constraints [32-34]. In particular, it was demonstrated that the existence of discrete breathers in nonlinear lattices with Hertzian contact between the oscillators without pre-compression is due to the presence of local elastic potentials in addition to the Hertzian interaction potentials [35]. In Section 2, the BA in a chain of linear damped oscillators with Hertzian coupling is demonstrated numerically. Detailed numeric exploration of the phenomenon in the space of parameters reveals main scaling relationships. In Section 3, we relate the breather arrest in the chain to the dynamics of beatings in a pair of linear damped oscillators with Hertzian coupling.  This latter system is simple enough to allow direct analytic exploration. Section 3 is followed by Conclusions.

## 2. Breather arrest phenomenon in a chain with Hertzian contact.

Let us consider a chain of *N* identical oscillators with mass *m* grounded by means of linear springs and viscous dampers with elastic stiffness and damping coefficient *k* and *d*, respectively. The oscillators in equilibrium are in contact without pre-compression, and interaction between them is described by generic function *g(z)*, as described below. The system is excited by application of instantaneous impact on the leftmost oscillator, after which the latter obtains initial velocity *V*, as shown in Figure 1.

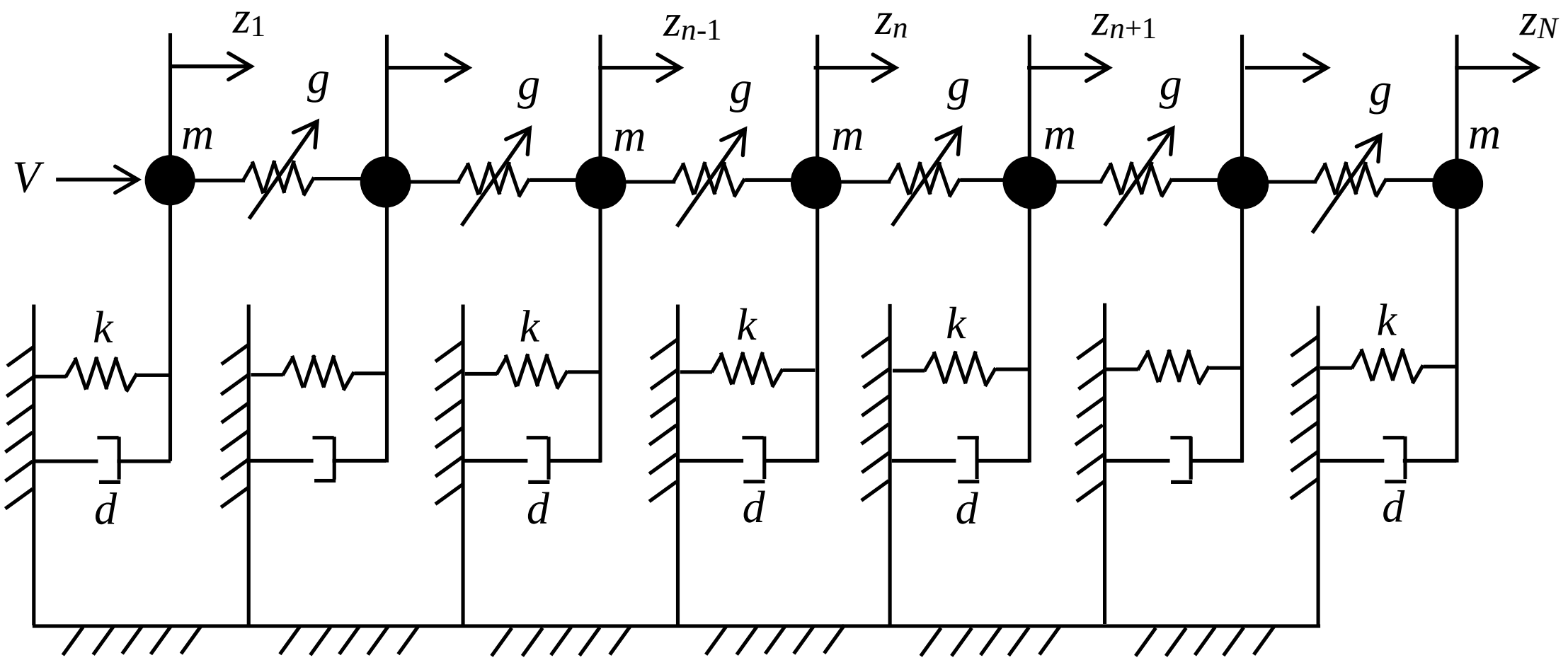

**Figure 1.** Schematic configuration of the dissipative oscillator chain with Hertzian contact.

The equation of motion for the *n*-th oscillator of the nonlinear chain of Figure 1 is written as:

$$mz_n''(t)+kz_n(t)+dz_n'(t)+Cg(z_n(t)-z_{n+1}(t))-Cg(z_{n-1}(t)-z_n(t))=0, \quad 1\le n\le N \tag{1}$$

where ($z_n, z_n', z_n''$) are the displacement, velocity and acceleration of the *n*-th oscillator, respectively, and *t* is the time, with $(\cdot)'=d(\cdot)/dt$, $C>0$ defines the coupling strength. In the case of Hertzian contact, the coupling function *g* between the nearest oscillators is given by:

$$g(y)=\begin{cases} y^{3/2}, & y>0 \\ 0, & y\le 0 \end{cases} \tag{2}$$

The form of function (2) implies that the Hertzian interactions between the oscillators are repulsive in presence of the contact (i.e. under compression, with distance reduction $y > 0$), while they vanish in absence of contact (i.e. under extension, with distance increase $y \le 0$), as unilateral constraints.

Starting from equation (1), the following set of $N$ ordinary differential equations of motion is found:

$$\begin{cases} mz_1''(t) + kz_1(t) + dz_1'(t) + Cg(z_1(t) - z_2(t)) = 0 \\ mz_2''(t) + kz_2(t) + dz_2'(t) + Cg(z_2(t) - z_3(t)) - Cg(z_1(t) - z_2(t)) = 0 \\ \cdots\cdots\cdots\cdots\cdots\cdots\cdots\cdots\cdots\cdots \\ mz_{N-1}''(t) + kz_{N-1}(t) + dz_{N-1}'(t) + Cg(z_{N-1}(t) - z_N(t)) - Cg(z_{N-2}(t) - z_{N-1}(t)) = 0 \\ mz_N''(t) + kz_N(t) + dz_N'(t) - Cg(z_{N-1}(t) - z_N(t)) = 0 \end{cases} \tag{3}$$

with initial conditions imposed for displacements and velocities in accordance with Figure 1:

$$\begin{cases} z_n(0) = 0, \quad 1 \le n \le N \\ z_n'(0) = 0, \quad 2 \le n \le N \\ z_1'(0) = V \end{cases} \tag{4}$$

By applying the following change of time and displacement variables:

$$t = \omega_n^{-1}\tau, \qquad z_n = \alpha u_n \tag{5}$$

where ($\tau$, $u_n$) are the dimensionless time and displacement of the $n$-th oscillator, respectively, $\omega_n$ is the circular frequency of the $n$-th oscillator and $\alpha$ is a reference displacement, and assuming:

$$k = m\omega_n^2, \qquad \lambda = \frac{d\omega_n}{k}, \qquad \alpha = \left(\frac{k}{C}\right)^2 > 0 \tag{6}$$

where $\lambda$ is the dimensionless damping coefficient, we can rewrite the equation of motion (1) for the $n$-th oscillator in the following dimensionless form:

$$\ddot{u}_n(\tau) + u_n(\tau) + \lambda\dot{u}_n(\tau) + f(u_n(\tau) - u_{n+1}(\tau)) - f(u_{n-1}(\tau) - u_n(\tau)) = 0, \qquad 1 \le n \le N \tag{7}$$

where ($\dot{u}_n, \ddot{u}_n$) define the dimensionless velocity and acceleration of the $n$-th oscillator, respectively, with $\dot{(\cdot)} = d(\cdot)/d\tau$.

The Hertzian nonlinear coupling function *g* between the nearest oscillators (2) also can be expressed in the dimensionless form:

$$f(x) = \begin{cases} x^{3/2}, & x > 0 \\ 0, & x \le 0 \end{cases} \tag{8}$$

Starting from equation (7), the following set of *N* dimensionless ordinary differential equations of motion with appropriate initial conditions is obtained:

$$\begin{cases} \ddot{u}_1(\tau) + u_1(\tau) + \lambda \dot{u}_1(\tau) + f(u_1(\tau) - u_2(\tau)) = 0 \\ \ddot{u}_2(\tau) + u_2(\tau) + \lambda \dot{u}_2(\tau) + f(u_2(\tau) - u_3(\tau)) - f(u_1(\tau) - u_2(\tau)) = 0 \\ \cdots\cdots\cdots\cdots\cdots\cdots\cdots\cdots\cdots\cdots\cdots\cdots\cdots\cdots \\ \ddot{u}_{N-1}(\tau) + u_{N-1}(\tau) + \lambda \dot{u}_{N-1}(\tau) + f(u_{N-1}(\tau) - u_N(\tau)) - f(u_{N-2}(\tau) - u_{N-1}(\tau)) = 0 \\ \ddot{u}_N(\tau) + u_N(\tau) + \lambda \dot{u}_N(\tau) - f(u_{N-1}(\tau) - u_N(\tau)) = 0 \end{cases}$$
$$\begin{cases} u_n(0) = 0, & 1 \le n \le N \\ \dot{u}_n(0) = 0, & 2 \le n \le N \\ \dot{u}_1(0) = A \end{cases} \tag{9}$$

Here $A = V/(\alpha\omega_1)$.

The presence of local elastic potentials in the nonlinear chain of Figure 1, with Hertzian interactions between the nearest oscillators and in absence of pre-compression, leads to the existence of discrete breathers [35]. In order to describe the breather arrest phenomenon, a direct numerical integration of the equations of motion (9) is performed by the *Wolfram Mathematica 9* computer software [36] using the Runge-Kutta iterative method. The following example is simulated for the parameters $N$ = 20, $A$ = 0.5, $\lambda$ = 0.1.

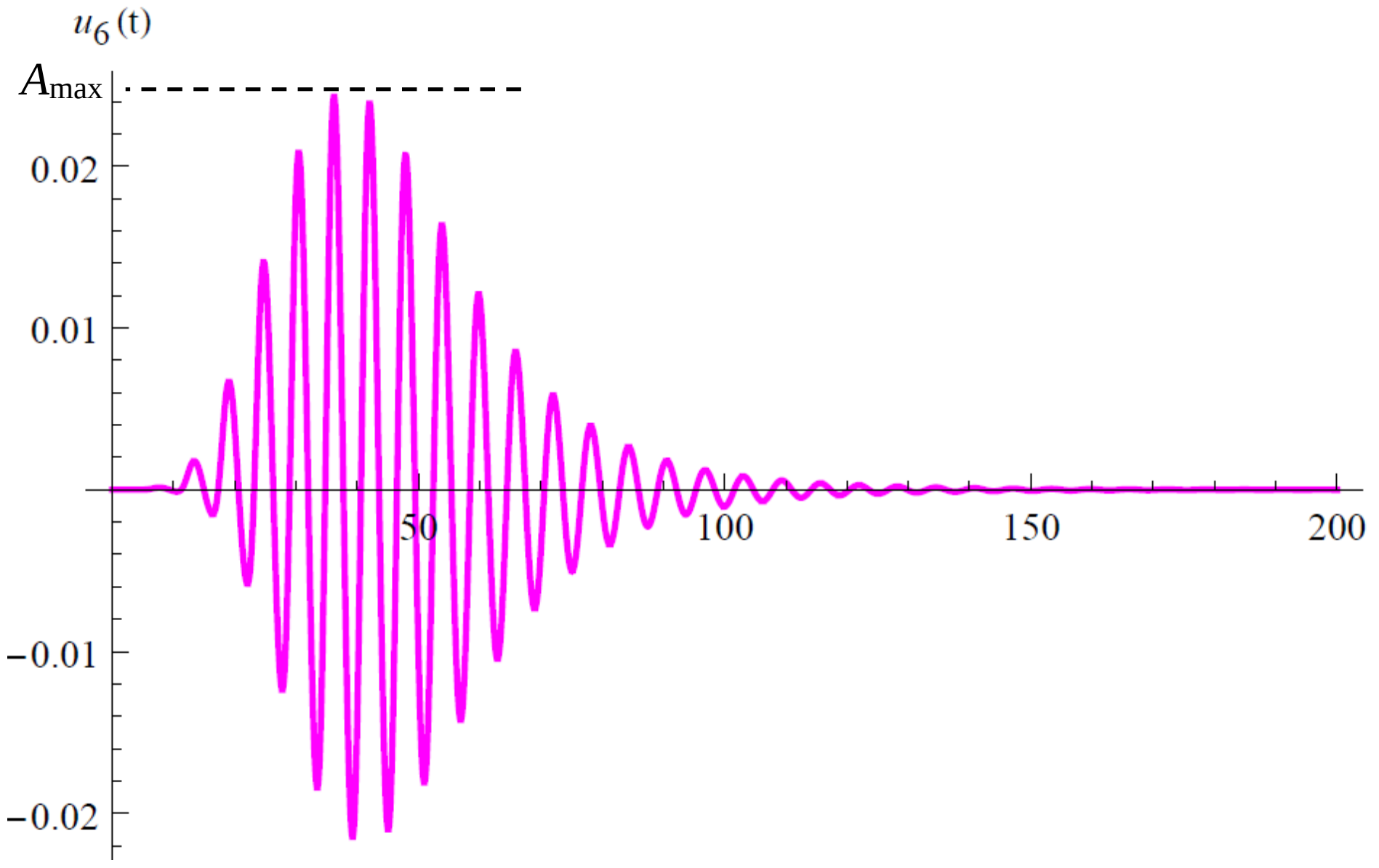

**Figure 2.** Nonlinear response of the oscillator $n$ = 6 from the left edge of the oscillatory chain of Figure 1. Numerical simulations with initial velocity amplitude $A$ = 0.5 and viscous damping coefficient $\lambda$ = 0.1.

As an example of the nonlinear response of one oscillator of the chain, in Figure 2 the displacement $u$ of the oscillator $n$ = 6 versus time is presented. The amplitude of displacement is very close to zero in the initial time region (no interactions with the nearest oscillators), then obtains maximum value $A_{max}$ in the central region with energy localization (where Hertzian interactions are present) and finally decreasing over the time $t$.

The travelling breather, represented by the set of nonlinear responses of the oscillators, propagates along the chain in the time starting from the first oscillator $n$ = 1 from the left edge subjected to the initial velocity amplitude $A$. The maximum amplitude $A_{max}$ of the travelling breather decreases from one oscillator to the next one along the chain due to the damping, as shown in Figure 3. The rate of decay of the travelling breather maximum amplitude along the chain depends on the values of viscous damping and initial velocity.

The BA is fixed numerically, when the maximum amplitude $A_{max}$ of the travelling breather on one oscillator of the chain assumes a value lower than the breather arrest threshold, which is selected as $A_{arrest} = 5\times10^{-4}$ in the present numerical analysis, in the form $A_{arrest} = A/1000$. The first oscillator from the left edge of the chain that satisfies the previous condition defines the “breather arrest depth” $n_{arrest}$. Of course, this number should be less than the length of the simulated chain. Intuitively, it is clear that for this sake the initial velocity should not be too high, and the damping – too low.

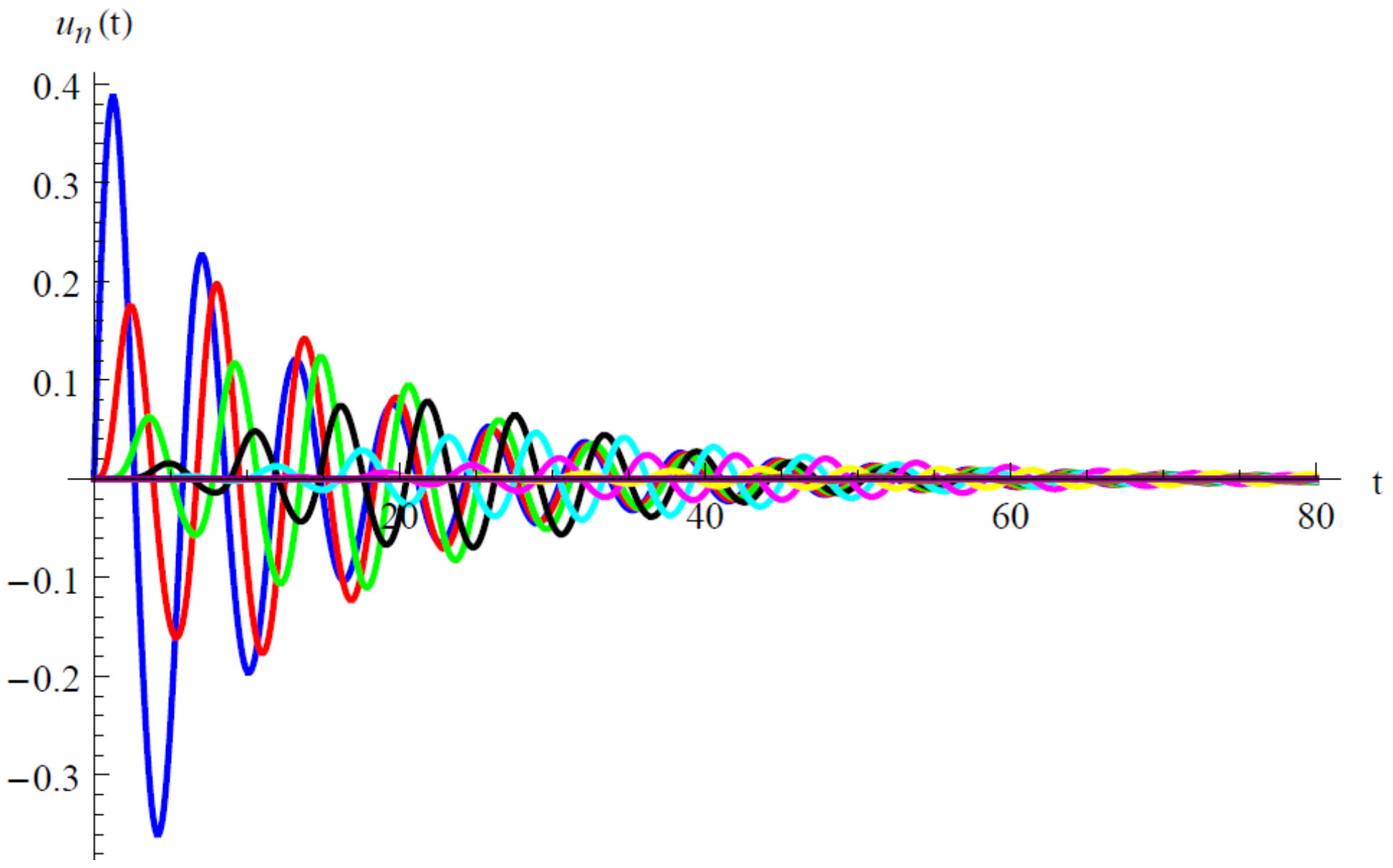

**Figure 3.** Travelling breather propagating along the oscillatory chain of Figure 1. Numerical simulations with initial velocity $A$ = 0.5 and damping coefficient $\lambda$ = 0.1. — $u_1(t)$, — $u_2(t)$, — $u_3(t)$, — $u_4(t)$, — $u_5(t)$, — $u_6(t)$, — $u_7(t)$, — $u_8(t)$.

Figure 4 depicts the maximum amplitude $A_{max}$ of the breather vs. the site index $n$. In this case, the breather arrest arises at the penetration depth $n_{arrest}$ = 9 where $A_{max} \approx 5\times10^{-4}$, which is equal to the BA threshold value $A_{arrest}$ previously selected.

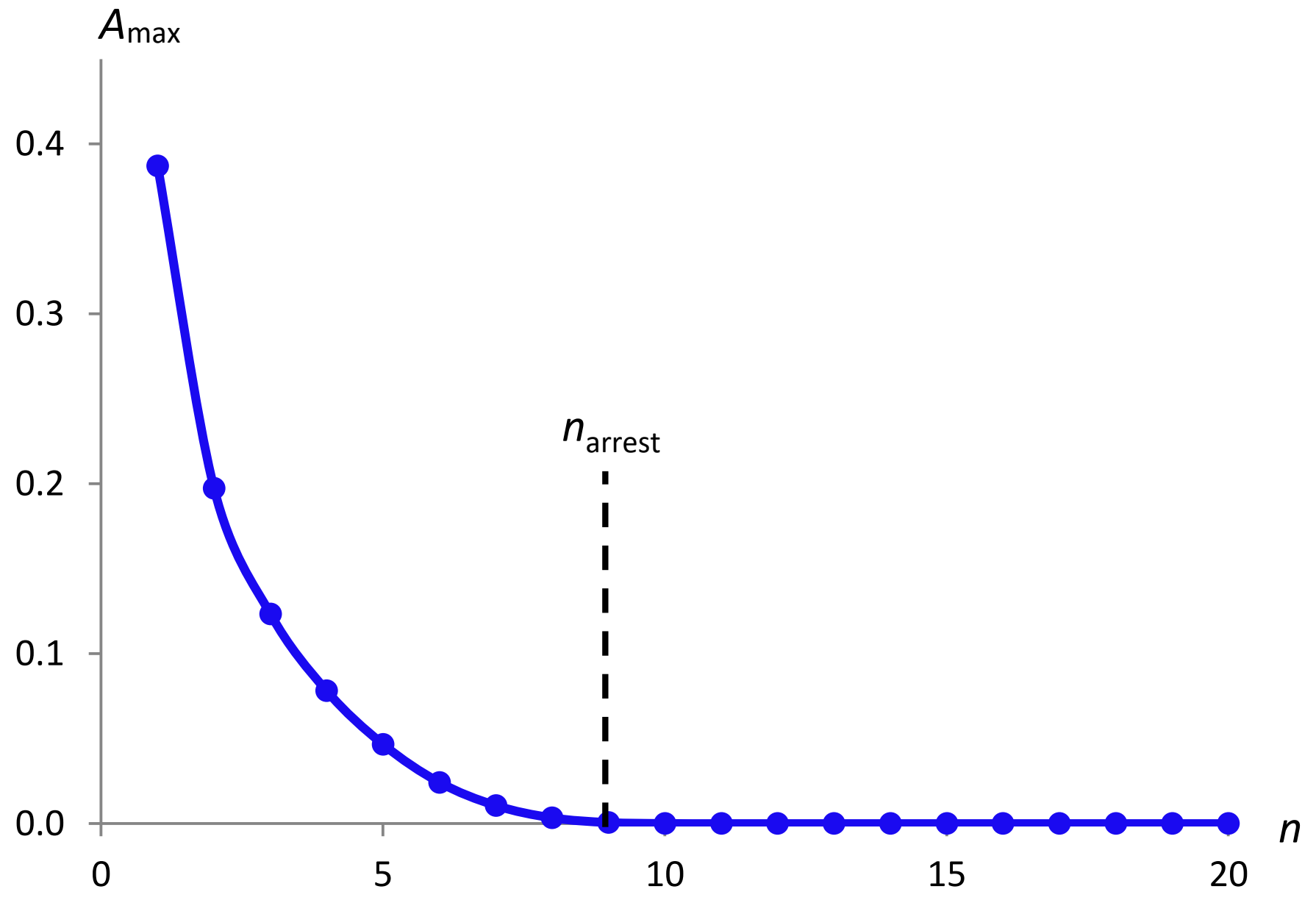

**Figure 4.** Maximum amplitude $A_{max}$ of the travelling breather vs. corresponding $n$-th oscillator along the chain of Figure 1. Numerical simulations with initial velocity $A$ = 0.5 and damping coefficient $\lambda$ = 0.1. Breather arrest depth $n_{arrest}$ = 9.

Figure 4 itself does nt provide sufficient information on the decay law. In Figure 5 we plot the logarithm of the breather maximum amplitude $A_{max}(n)$ versus the logarithm of the quantity ($n_{arrest}$ – $n$), with the BA depth value $n_{arrest} = 9$. This figure implies linear dependence between $\ln(A_{max}(n))$ and $\ln(n_{arrest} - n)$ close to the arrest, with approximate slope $s = 2.08$.

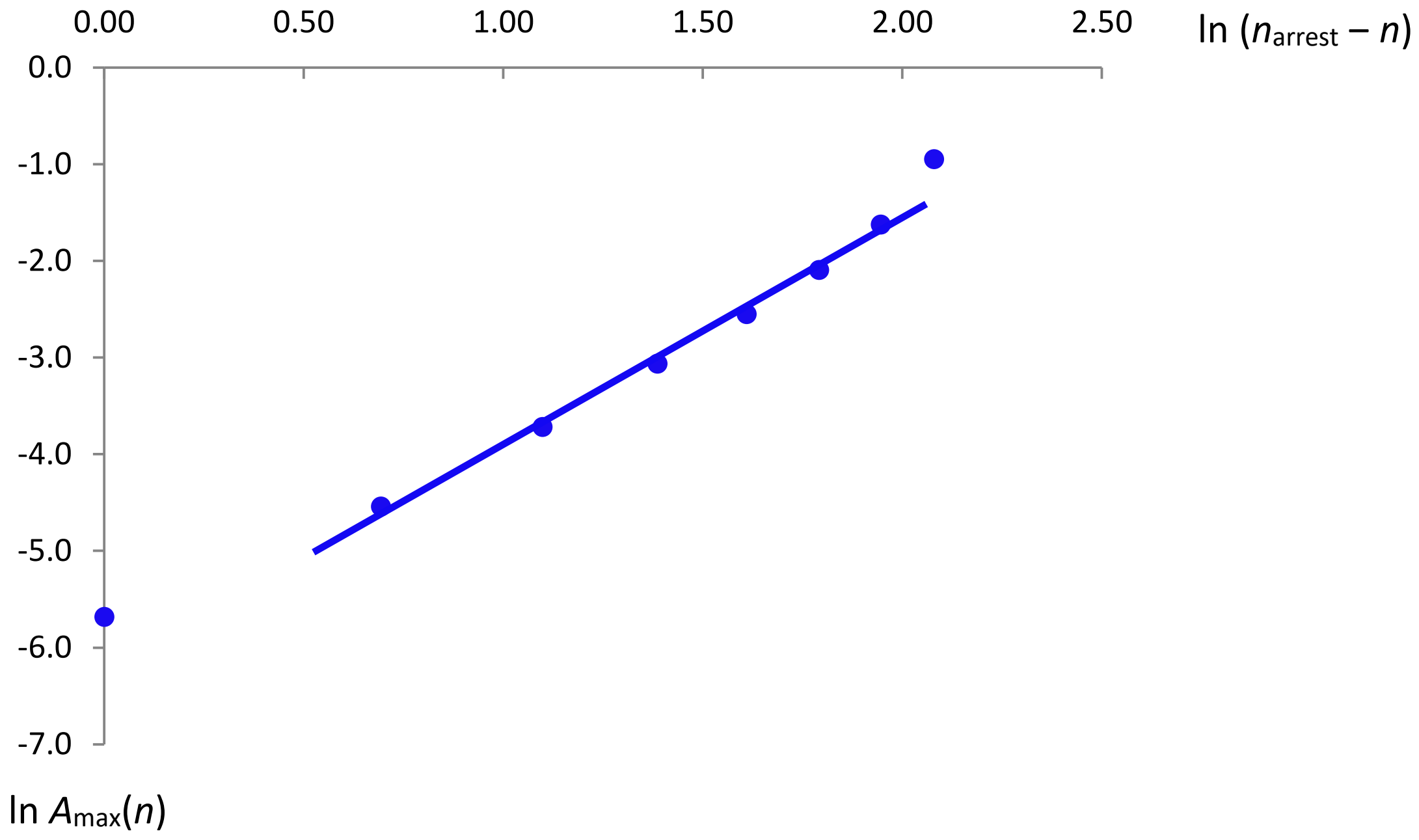

**Figure 5.** Breather maximum amplitude $A_{max}(n)$ vs. $(n_{arrest} - n)$ in logarithmic coordinates. Initial velocity $A = 0.5$ and damping coefficient $\lambda = 0.1$. Breather arrest depth $n_{arrest} = 9$.

Therefore, one can conjecture that the decay of the breather amplitude at the initial stage of its propagation is described as:

$$A_{max}(n) \sim (n_{arrest} - n)^{s} \tag{10}$$

Definitely, estimation (10) cannot hold for $n > n_{arrest}$. From the other side, the particles in the chain are connected, and therefore the excitation will inevitably pass along the whole chain. In order to investigate the rate of decay of the breather maximum amplitude along the whole chain, in Figure 6 we plot the logarithm of the breather maximum amplitude $A_{max}$ vs. the corresponding oscillator $n$. One clearly observes that for $n > n_{arrest}$ the breather decays at hyper-exponential rate. Thus, we observe that the breather propagation along the chain is clearly divided into two stages with qualitatively different asymptotic behaviour. Thus, the notion of the arrest is justified, due to sharp transformation in the propagation pattern – from the power-law to the hyper-exponential decay.

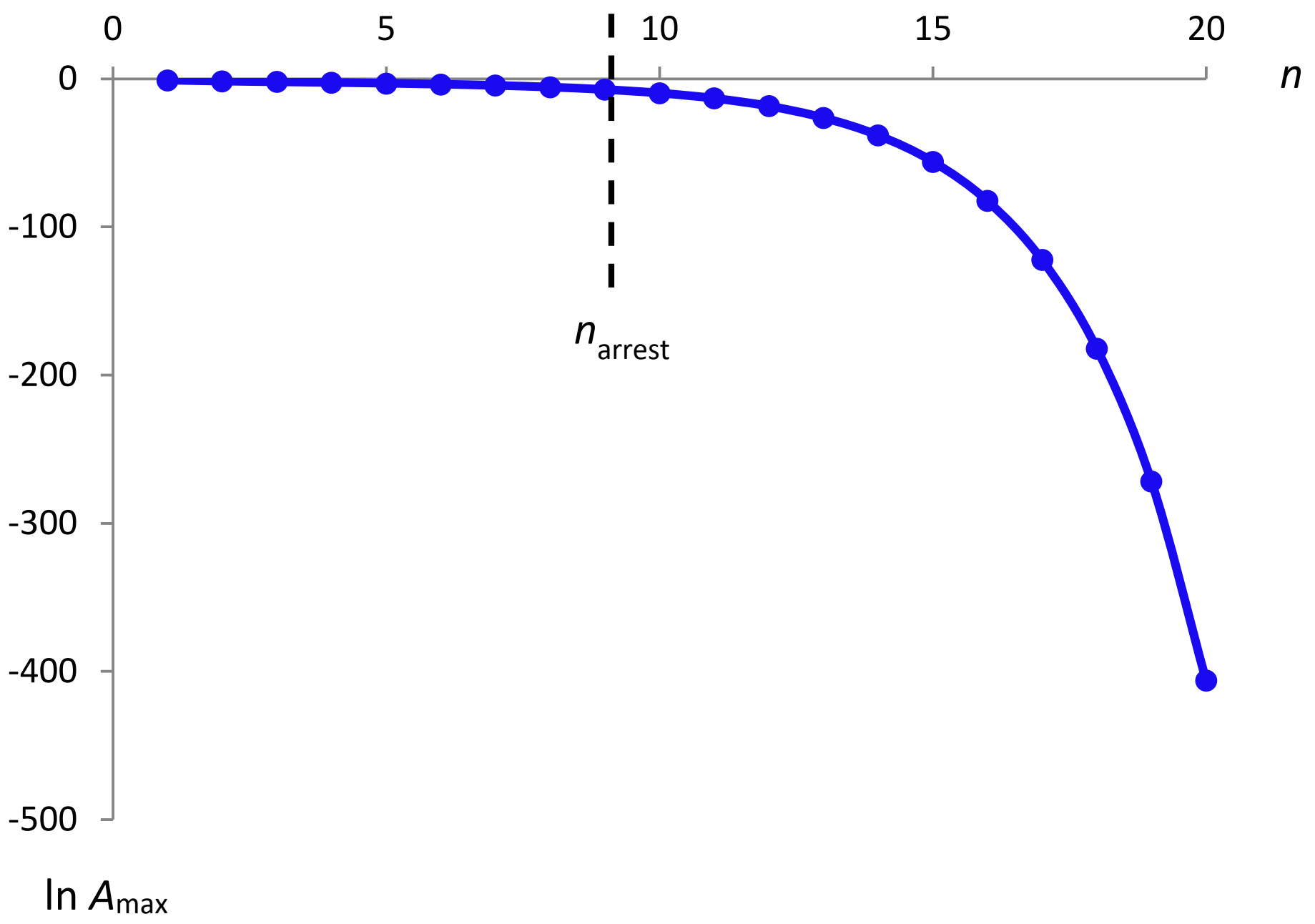

**Figure 6.** Logarithm of the maximum amplitude $A_{max}$ of the travelling breather vs. corresponding *n*-th oscillator along the nonlinear chain of Figure 1. Numerical simulations with initial velocity $A = 0.5$ and damping coefficient $\lambda = 0.1$.

## 2.1. Numerical results

In this section, we present more detailed numeric simulations of the BA phenomenon in the space of parameters. In Figure 7, dependencies of the BA depth $n_{arrest}$ on the initial velocities in the range $A$ = 0.000 ÷ 0.010 for a set of values of the damping coefficient are plotted. From these curves, it can be observed that, for a constant viscous damping coefficient, the breather arrest depth increases with increasing the initial velocity amplitude; besides, for a constant initial velocity amplitude, the BA depth increases while the damping decreases.

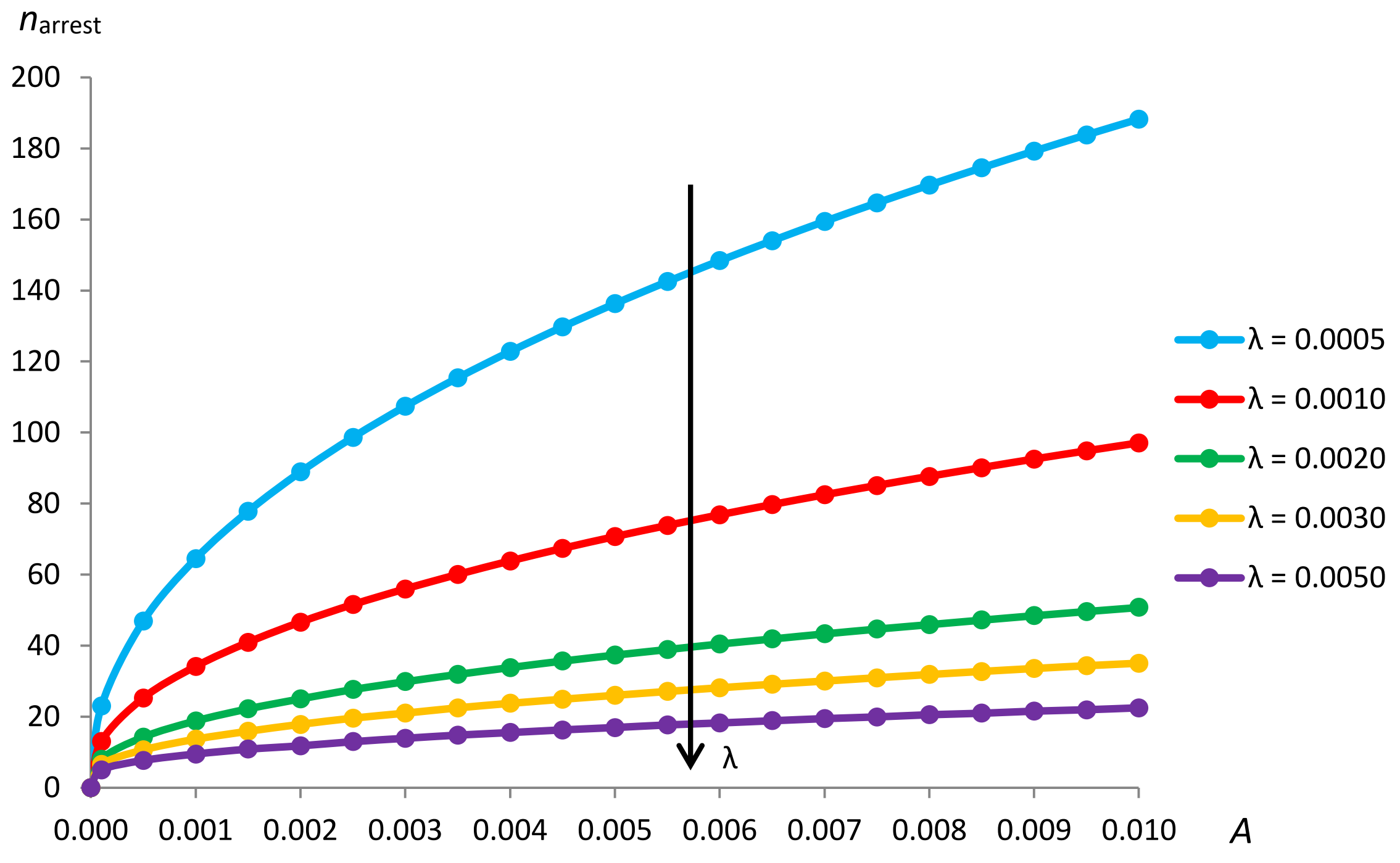

**Figure 7.** Curves of BA depth $n_{arrest}$ vs. initial velocity $A$ for different damping coefficients $\lambda$.

The curves of Figure 7 are plotted in logarithmic coordinates in Figure 8 for the initial velocity range $A$ = 0.008 ÷ 0.010; the angular coefficients of the corresponding straight line – approximations are reported in Table 1.

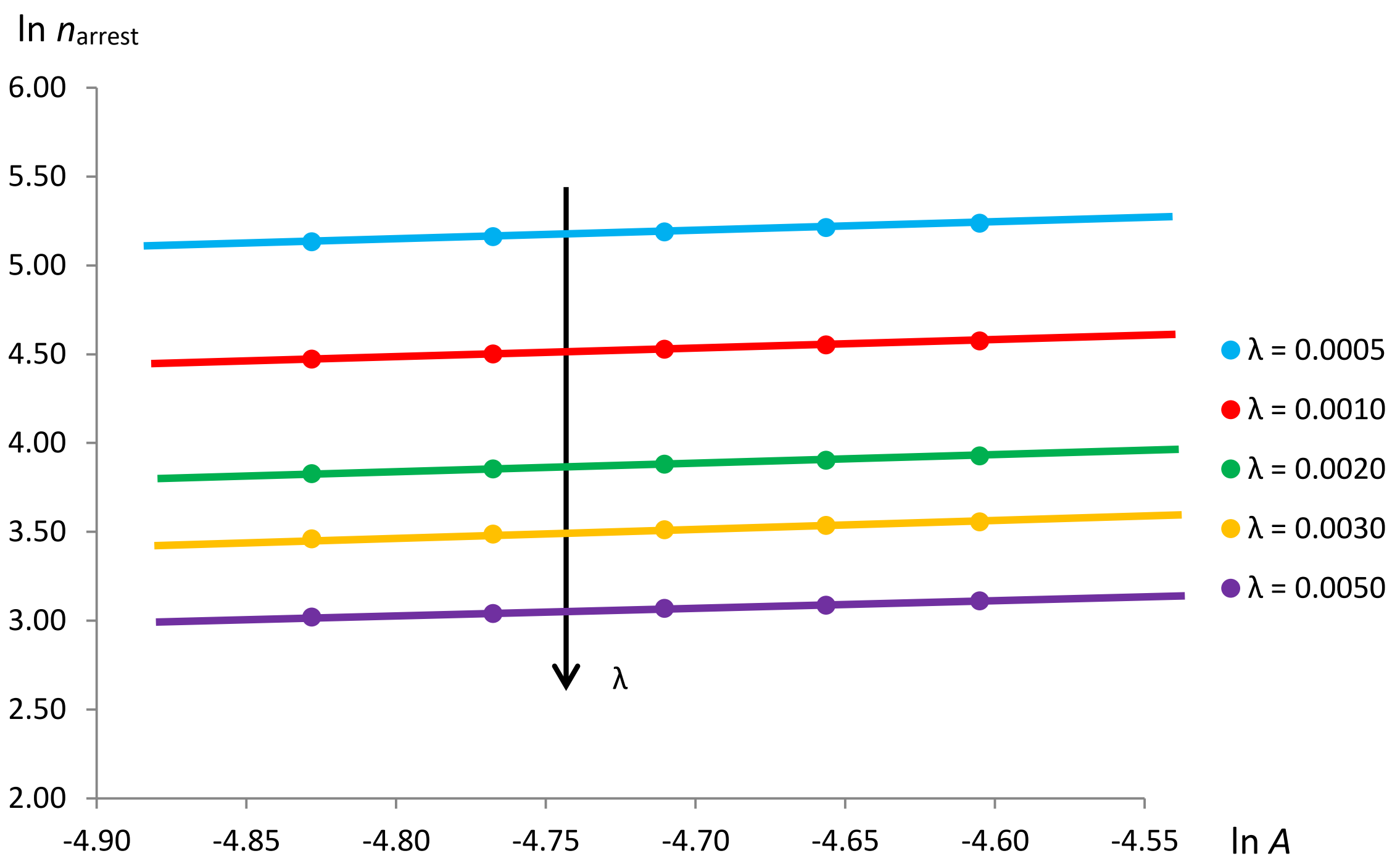

**Figure 8.** Curves of BA depth $n_{arrest}$ vs. initial velocity $A$ in logarithmic coordinates for different damping coefficients $\lambda$. Initial velocity range $A$ = 0.008 ÷ 0.010.

| Damping coefficient $\lambda$ | Angular coefficient $q$ |
|---|---|
| 0.0005 | 0.466 |
| 0.0010 | 0.457 |
| 0.0020 | 0.448 |
| 0.0030 | 0.428 |
| 0.0050 | 0.407 |

**Table 1.** Angular coefficients $q$ of the straight curves in logarithmic coordinates of Figure 8 for different values of damping coefficient $\lambda$. Initial velocity range $A$ = 0.008 ÷ 0.010.

From Table 1, it is noted that the angular coefficient of the straight lines of Figure 8 increases with decreasing the damping coefficient $\lambda$, where the maximum angular coefficient $q =$ 0.466 is found at the minimum damping coefficient $\lambda$ = 0.0005.

In Figure 9, the dependencies of $n_{arrest}$ on the damping coefficient in a range $\lambda$ = 0.000 ÷ 0.010 for different values of the initial velocity are presented. From these curves, similarly to the curves of Figure 7, one can note that, for a constant initial velocity amplitude, the breather arrest depth increases decreasing the viscous damping coefficient; on the other hand, for a constant viscous damping coefficient, the breather arrest depth increases increasing the initial velocity amplitude.

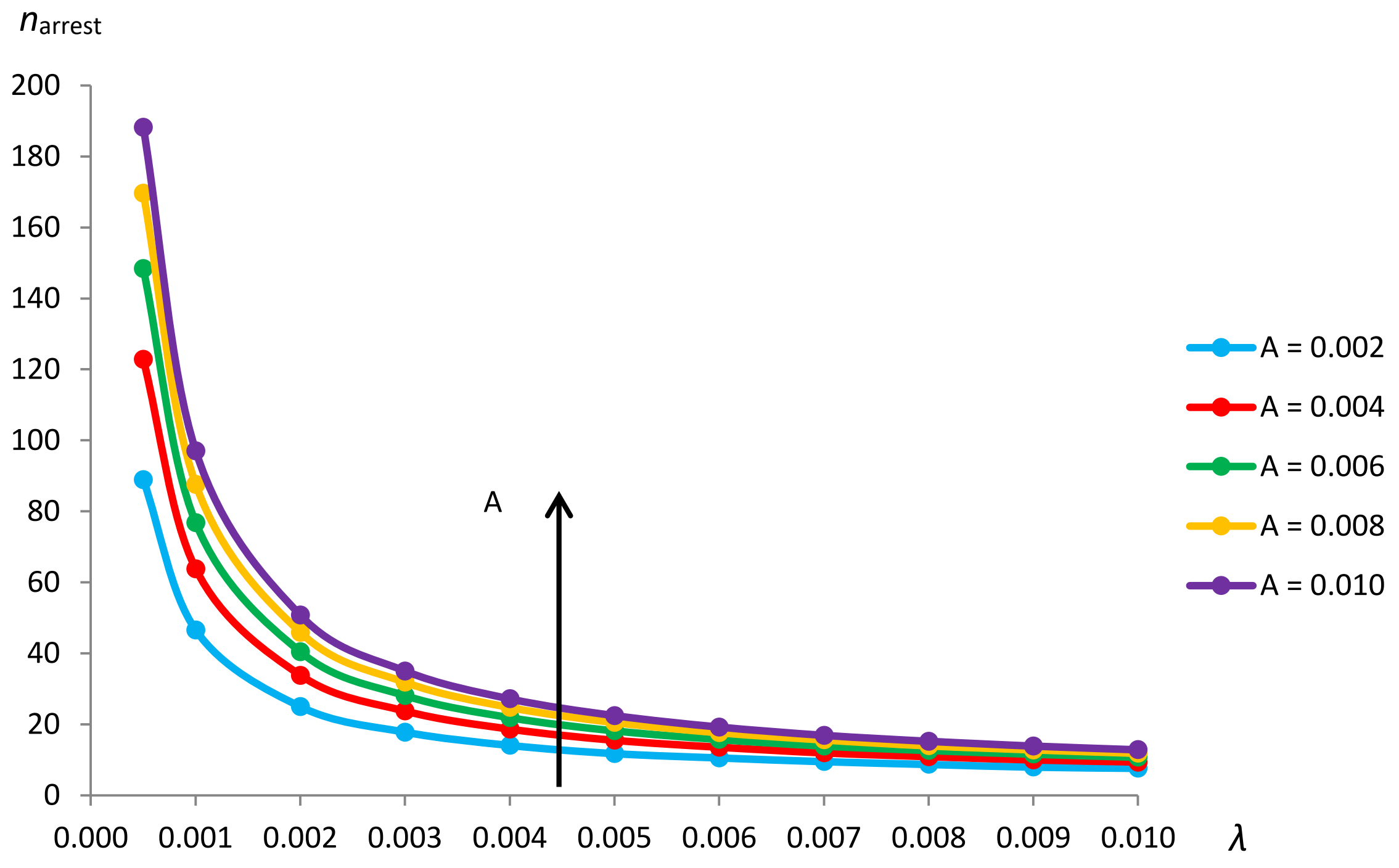

**Figure 9.** Curves of BA depth $n_{arrest}$ vs. damping coefficient $\lambda$ for different initial velocities $A$.

The curves from Figure 9 are shown for the damping coefficient range λ = 0.0005 ÷ 0.0040 in logarithmic coordinates in Figure 10 and the angular coefficients of the corresponding straight line – approximations are listed in Table 2.

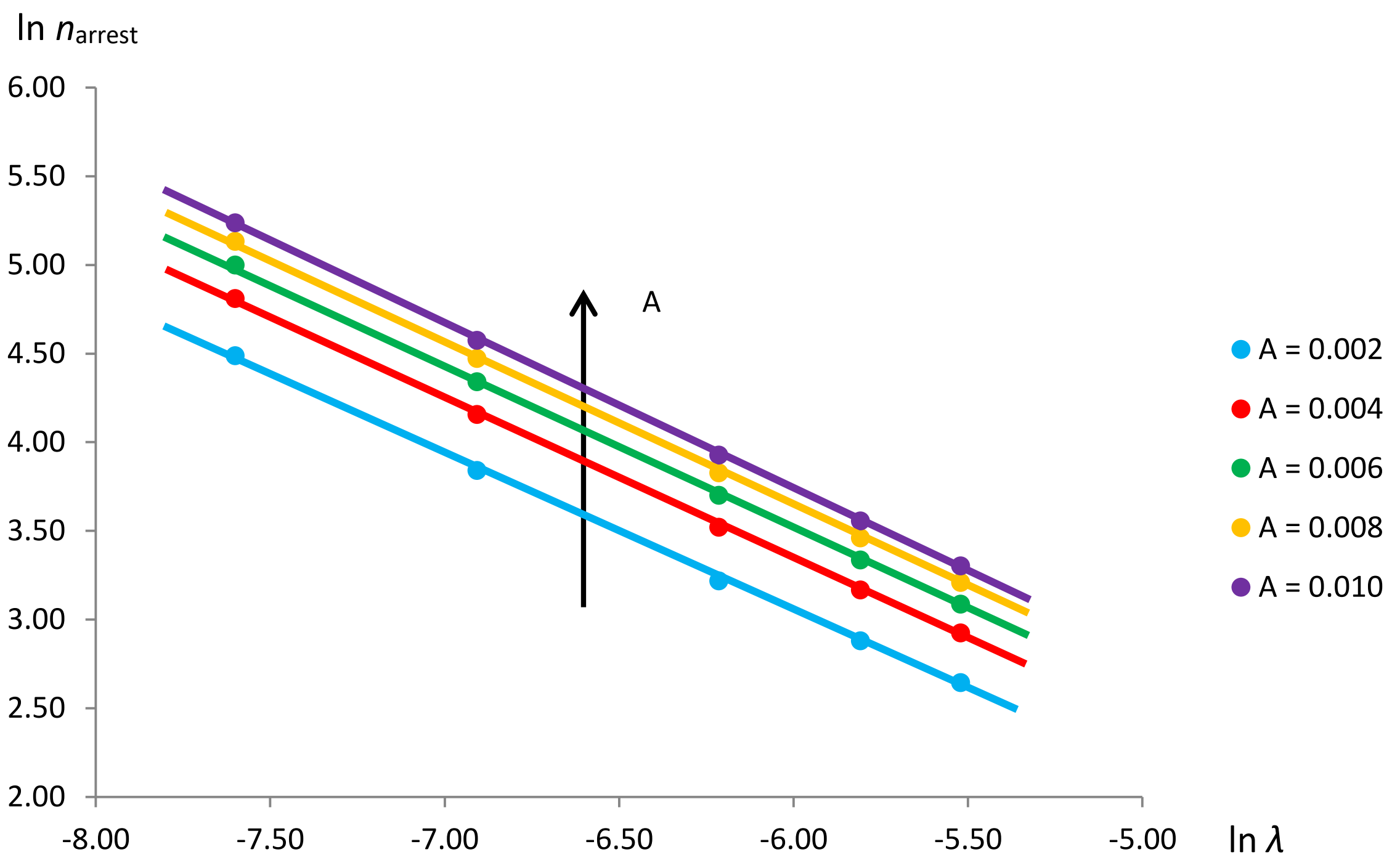

**Figure 10.** Curves of BA depth $n_{arrest}$ vs. damping coefficient $\lambda$ in logarithmic coordinates for different initial velocities $A$. Damping coefficient range $\lambda = 0.0005 \div 0.0040$.

| Initial velocity $A$ | Angular coefficient $l$ |
|---|---|
| 0.002 | − 0.886 |
| 0.004 | − 0.907 |
| 0.006 | − 0.920 |
| 0.008 | − 0.926 |
| 0.010 | − 0.931 |

**Table 2.** Angular coefficient $l$ of the straight curves in logarithmic coordinates of Figure 10 for different values of initial velocity $A$. Damping coefficient range $\lambda = 0.0005 \div 0.0040$.

From Table 2, it is seen that the angular coefficient of the straight lines of Figure 10 decreases with increasing the initial velocity $A$, where the minimum angular coefficient $l = -0.931$ is found at the maximum initial velocity $A = 0.010$.

In view of the simulation results, is natural to assume that the breather arrest depth scales with the initial velocity and the damping coefficient according to the approximate formula:

$$n_{\mathrm{arrest}} \sim A^{q}\lambda^{l} \qquad (11)$$

Numeric values of the scaling exponents $q$ and $l$ are presented in Tables 1 and 2 respectively.

## 3. The simplified model of two oscillators

To assess the BA phenomenon from theoretical viewpoint, and to rationalize the numeric findings presented in the previous Section, we suggest a simplified model that mimics the breather propagation in the chain with strongly nonlinear coupling. This simplification seems viable due to extreme localization of the breather in this chain, well-known for the chains with strongly nonlinear interactions [32,33,35]. Therefore, it is possible to adopt, in the crudest approximation, that the breather propagation can be understood as a sequence of energy transfers between subsequent particles. Moreover, due to the strong localization, it is possible to assume that each such transfer involves only two neighbouring particles. Thus, the simplified model consists of two identical oscillators with mass *m*, grounded through pairs of linear springs and viscous dampers with elastic stiffness and damping coefficient *k* and *d*, respectively. These oscillators are coupled by some strongly nonlinear (nonlinearizable) spring. Initially, the nonzero velocity *V* is assumed at the left oscillator, as shown in Figure 11. Then, the beating motion occurs. Each beating event is associated with propagation of the breather by one particle. We are going to demonstrate that the number of such beatings will be finite, due to the strong nonlinearity of the coupling.

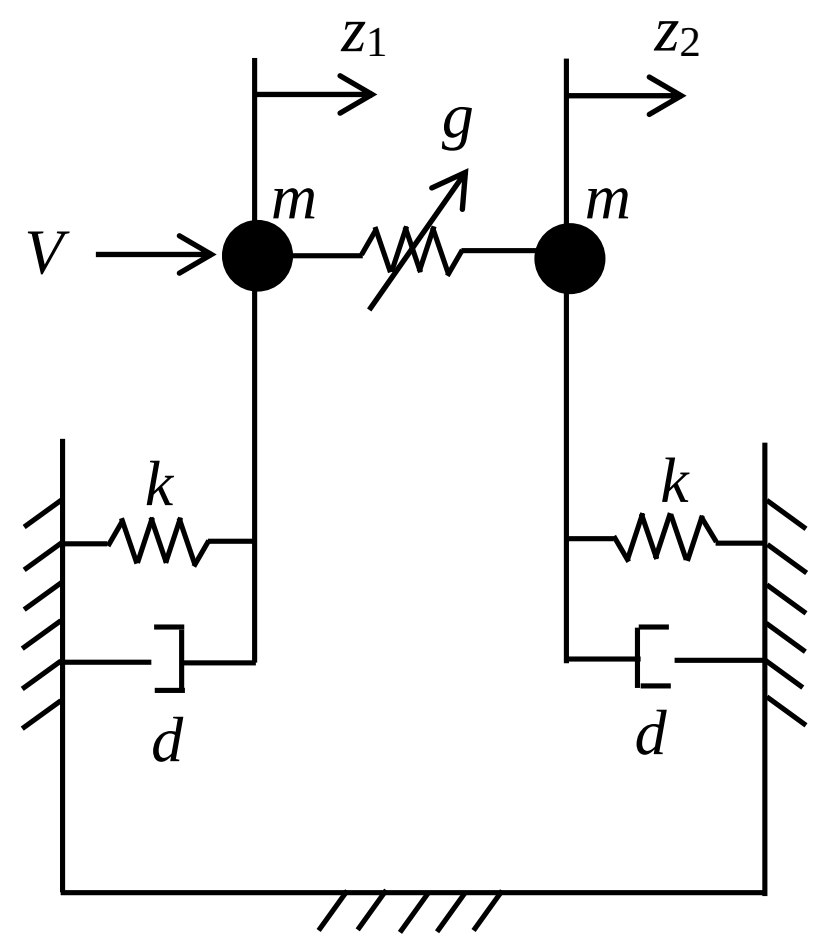

**Figure 11.** Schematic configuration of the two-coupled oscillators with Hertzian contact.

By considering the relations (5-6), the equations of motion for the two-coupled oscillators of Figure 11 can be written in dimensionless form as follows (2DOF approximation):

$$\begin{cases} \ddot{u}_1(\tau) + u_1(\tau) + \lambda \dot{u}_1(\tau) + f(u_1(\tau) - u_2(\tau)) = 0 \\ \ddot{u}_2(\tau) + u_2(\tau) + \lambda \dot{u}_2(\tau) - f(u_1(\tau) - u_2(\tau)) = 0 \end{cases} \tag{12}$$

where $(u_i, \dot{u}_i, \ddot{u}_i)$ are the dimensionless displacement, velocity and acceleration of the *i*-th oscillator, respectively, with $i$ = (1,2), $\lambda$ is the dimensionless damping coefficient and the corresponding dimensionless initial conditions on displacements and velocities are:

$$\begin{cases} u_1(0)=0 \\ u_2(0)=0 \end{cases} \qquad \begin{cases} \dot{u}_1(0)=A \\ \dot{u}_2(0)=0 \end{cases} \tag{13}$$

The coupled 2-dof system (12) can be uncoupled by introducing the following dependent variables:

$$\begin{cases} R=\dfrac{1}{2}(u_1+u_2) \\ w=u_1-u_2 \end{cases} \tag{14}$$

where $R$ is the centre of masses and $w$ is the internal displacement of the system, in the form:

$$\begin{cases} \ddot{R}+R+\lambda\dot{R}=0 \\ \ddot{w}+w+\lambda\dot{w}+2f(w)=0 \end{cases} \tag{15}$$

and the initial conditions (13) assume the following form:

$$\begin{cases} R(0)=0 \\ w(0)=0 \end{cases} \qquad \begin{cases} \dot{R}(0)=A/2 \\ \dot{w}(0)=A \end{cases} \tag{16}$$

In what follows, we assume that the damping is relatively small (to ensure substantial breather propagation, or number of beatings).

Then, in the basic approximation, while neglecting the high-order corrections, the solution for the centre of masses $R$ is written as:

$$R(\tau)\approx\frac{A}{2}\exp\left(-\frac{\lambda\tau}{2}\right)\sin\tau \tag{17}$$

As for the internal displacement $w$, we have to find the approximate solution for the following problem:

$$\begin{cases} \ddot{w}(\tau) + w(\tau) + \lambda \dot{w}(\tau) + 2f(w(\tau)) = 0 \\ w(0) = 0 \\ \dot{w}(0) = A \end{cases} \tag{18}$$

In order to analyse this problem, we first perform the transformation to action-angle variables [37] generated by the Hamiltonian:

$$H = \frac{p^2}{2} + \frac{w^2}{2} + 2F(w), \quad p = \dot{w}, \quad F'(w) = f(w), \quad F(0) = 0 \tag{19}$$

Dependence between energy of the undamped motion and the action is determined by well-known formula [37]:

$$2\pi I = \oint p dw = 2 \int_{w_{\min}}^{w_{\max}} \sqrt{2E - w^2 - 4F(w)} dw, \quad \theta = \frac{\partial}{\partial I} \int_0^{w(I,\theta)} \sqrt{2E(I) - w^2 - 4F(w)} dw \tag{20}$$

where $E$ is the energy of the undamped motion.

This canonical transformation $(p,w) \to (I,\theta)$ brings Equation (18) to the following form:

$$\dot{I} = -\lambda p(I,\theta) \frac{\partial q(I,\theta)}{\partial \theta}, \quad \dot{\theta} = \lambda p(I,\theta) \frac{\partial q(I,\theta)}{\partial I} + \frac{\partial E(I)}{\partial I} \tag{21}$$

Assuming the weak damping, one can perform the primary averaging of system (21), to obtain:

$$\dot{J} = -\frac{\lambda}{2\pi} \oint p dq = -\lambda J, \quad \langle \dot{\theta} \rangle = \frac{\partial E(J)}{\partial J}, \quad J = \langle I \rangle_\theta \tag{22}$$

The term $p\ (\partial q / \partial I)$ in the second equation of system (21) vanishes by primary averaging, due to the invariance of the Hamiltonian (19) with respect to $p \to -p$, as it is always possible to choose the transformations to ensure the conditions:

$$q(I,\theta) = q(I,-\theta) \Rightarrow \frac{\partial q(I,\theta)}{\partial I} = \frac{\partial q(I,-\theta)}{\partial I}, \quad p(I,\theta) = -p(I,-\theta) \tag{23}$$

Equations (22) are easily solvable and yield:

$$J = J_0 \exp(-\lambda\tau), \qquad \langle\theta\rangle = \theta_0 + \int_0^\tau \Omega(J_0 \exp(-\lambda\tau))d\tau, \qquad \Omega = \frac{\partial E(J)}{\partial J} \tag{24}$$

When discussing the breather arrest, we are primarily interested in terminal stage of the beatings in system (12), when it is possible to adopt in the basic approximation that the motion is already quasi-linear, and to use the well-known action-angle transformation for linear oscillator [37]:

$$w \approx \sqrt{2J(\tau)}\sin\langle\theta\rangle \approx A\exp\left(-\frac{\lambda\tau}{2}\right)\sin\left(\int_0^t \Omega\left(\frac{A^2}{2}\exp(-\lambda\tau)\right)d\tau\right) \tag{25}$$

Still, estimation of the instantaneous frequency $\Omega(J)$ requires more refined analysis. Again, assuming small amplitudes in equation (20), one obtains:

$$E = I + \frac{2}{\pi}\int_{-1}^{1}\frac{F(\sqrt{2I}w)}{\sqrt{1-w^2}}dw + \text{H.O.T.} \tag{26}$$

Here H.O.T. stands for high-order terms. Equation (26) is general and suitable for any nonlinear coupling, provided that the integral is not zero. If the integral in Equation (26) turns out to be zero, these high-order terms should be analysed more thoroughly, since they provide the primary effect. However, in the case of Hertzian contact considered in the paper it is not the case. For the specific case of Hertzian contact, one adopts:

$$f(w) = \begin{cases} 0, & w \le 0 \\ w^{3/2}, & w > 0 \end{cases} \quad \Rightarrow \quad F(w) = \begin{cases} 0, & w \le 0 \\ \frac{2}{5}w^{5/2}, & w > 0 \end{cases} \tag{27}$$

Substituting (27) into (26), one obtains:

$$E = I + \frac{24}{25\pi}2^{3/4}I^{5/4}\left(2\mathbf{E}\left(\frac{\sqrt{2}}{2}\right) - \mathbf{K}\left(\frac{\sqrt{2}}{2}\right)\right) \tag{28}$$

Here $\mathbf{K}(k)$, $\mathbf{E}(k)$ are complete elliptic integrals of the first and the second kind, respectively.
After simple algebra, one obtains:

$$w(\tau) \approx A\exp\left(-\frac{\lambda\tau}{2}\right)\sin\left(\tau + 2c\frac{A^{1/2}}{\lambda}\left(1-\exp\left(-\frac{\lambda t}{4}\right)\right)\right), \quad c = \frac{12\sqrt{2}}{5\pi}\left(2\mathbf{E}\left(\frac{\sqrt{2}}{2}\right)-\mathbf{K}\left(\frac{\sqrt{2}}{2}\right)\right) \tag{29}$$

Finally, for the initial variables of the problem $u_1$, $u_2$ one obtains:

$$\begin{aligned} u_1(\tau) &= R(\tau) + \frac{w(\tau)}{2} \approx \\ &\approx A\exp\left(-\frac{\lambda\tau}{2}\right)\sin\left(\tau + c\frac{A^{1/2}}{\lambda}\left(1-\exp\left(-\frac{\lambda t}{4}\right)\right)\right)\cos\left(c\frac{A^{1/2}}{\lambda}\left(1-\exp\left(-\frac{\lambda t}{4}\right)\right)\right), \\ u_2(\tau) &= R(\tau) - \frac{w(\tau)}{2} \approx \\ &\approx A\exp\left(-\frac{\lambda\tau}{2}\right)\cos\left(\tau + c\frac{A^{1/2}}{\lambda}\left(1-\exp\left(-\frac{\lambda t}{4}\right)\right)\right)\sin\left(c\frac{A^{1/2}}{\lambda}\left(1-\exp\left(-\frac{\lambda t}{4}\right)\right)\right) \end{aligned} \tag{30}$$

The number of beatings is determined by the last terms in equations (30). This number (i.e., the breather penetration depth) is determined as:

$$n_{\max} \approx n_{\text{arrest}} \sim \frac{2c}{\pi}\frac{A^{1/2}}{\lambda} \tag{31}$$

Equation (31) explains the power laws numerically stated above in Equation (11). The theoretical values of the scaling exponents $q = 1/2$, $l = -1$ are in good agreement with numeric findings listed in Tables 1,2 in the previous Section.
To explain the results in Figures 4-5 and approximation (10) concerning the decay of the breather $A_{\max}$ as a function of the site number $n$, we note that the maxima are approximately achieved at time instances when the last terms in (30) achieve zero or extrema. Then, the maximum amplitude $A_{\max}$ at the site $n$ is achieved at the time instance $\tau_n$, determined by the following relationship:

$$n_{\text{arrest}}\left(1-\exp\left(-\frac{\lambda\tau_n}{4}\right)\right) \approx n \tag{32}$$

Combining Equation (32) with Equation (30), one obtains the following estimation for the breather maximum amplitude:

$$A_{\max}(n) \approx A\left(1-\frac{n}{n_{\text{arrest}}}\right)^2 \tag{33}$$

Again, this expression nicely fits to numeric findings of the previous Section, see Figure 5 and Equation (10). As one can expect, Equation (33) is valid only for finite number of sites, and does not explain the hyper-exponential propagation process shown in Figure 6. This process is obviously beyond the validity of the simplified model pursued in this Section.

Finally, let us consider more general power law in Equation (27):

$$f(w) = \begin{cases} 0, & w \le 0 \\ w^{\delta}, & w > 0 \end{cases}, \quad \delta > 1 \tag{34}$$

Then, by similar evaluation, one obtains the following general scaling laws:

$$n_{\text{arrest}} \sim \frac{A^{\delta-1}}{\lambda}, \quad A_{\max}(n) \approx A\left(1-\frac{n}{n_{\text{arrest}}}\right)^{\frac{1}{\delta-1}} \tag{35}$$

Earlier paper [31] has addressed the case $\delta = 3$(the potential of interaction was symmetric, but this peculiarity has no effect on scaling estimations like (35)). Numeric exploration indeed confirmed the dependence $n_{arrest} \sim A^2/\lambda$, but the breather amplitude decay has not been explored.

To supply additional illustration for the scaling laws (35), we present in Figure 12 the dependence of the breather amplitude $A_{\max}$ on the site $n$ for the case of purely cubic coupling, i.e. for $f(w) = w^3$. Other parameters are listed in the figure caption. In this figure, one again observes the two stages of the breather propagation – pre-arrest and after-arrest, with even more clear distinction between them – the plot inflects near the arrest point.

Figure 13 shows the breather propagation in double-logarithmic coordinates: since the relationship between $A_{\max}$ and ($n_{\text{arrest}} - n$) close to $n_{\text{arrest}}$ is approximately represented by a straight line with slope $s$ = 0.51, then this figure corroborates the estimation $A_{\max}(n) \sim \sqrt{n_{\text{arrest}} - n}$ for the pre-arrest propagation that follows from estimation (35) for $\delta = 3$.

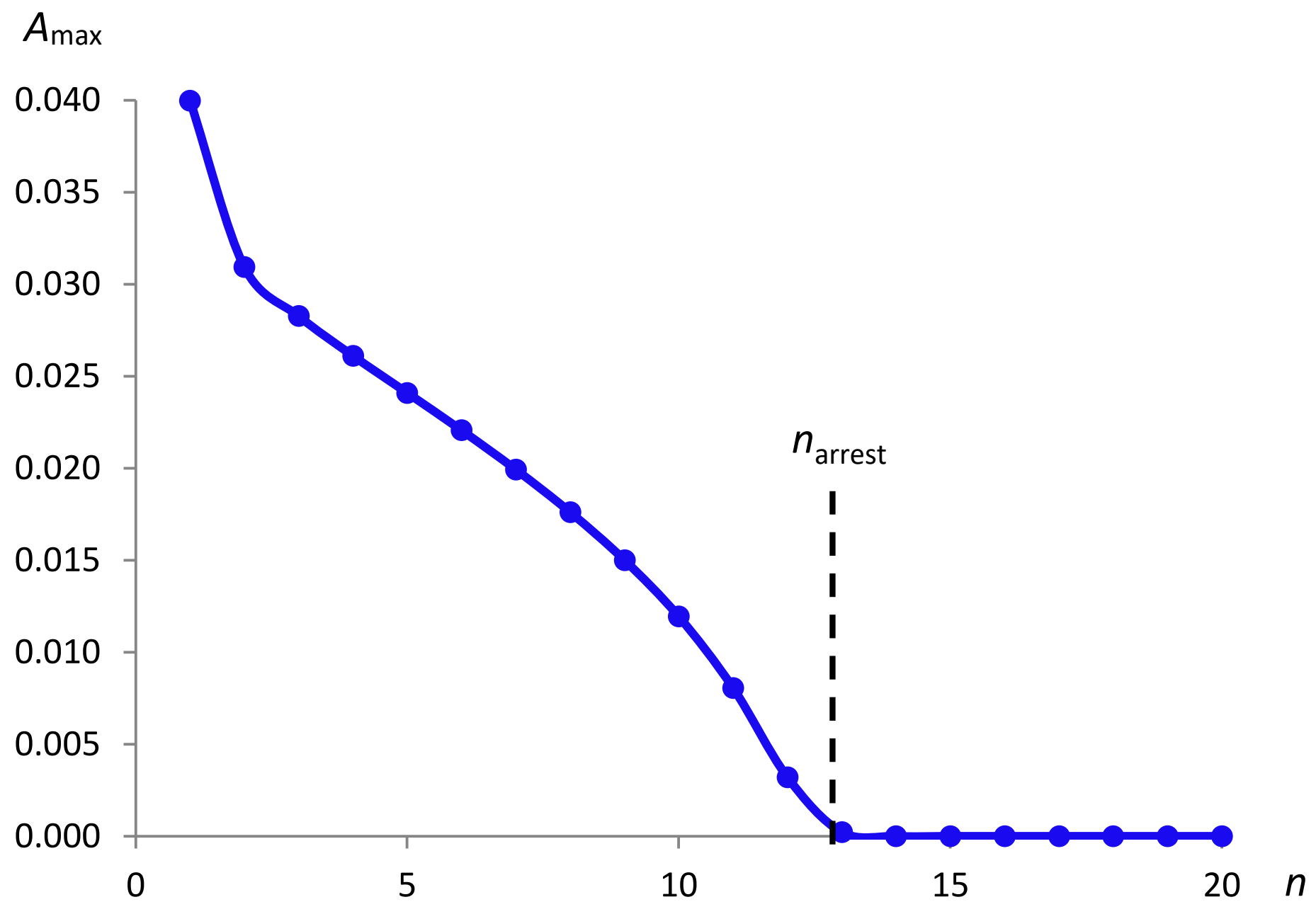

**Figure 12.** Maximum amplitude $A_{max}$ of the travelling breather vs. corresponding oscillator $n$ along a nonlinear chain of damped linear oscillators with cubic coupling between the nearest neighbours. Numerical simulations with initial velocity amplitude $A = 0.04$ and viscous damping coefficient $\lambda = 0.00005$. Travelling breather arrest depth $n_{arrest} = 13$.

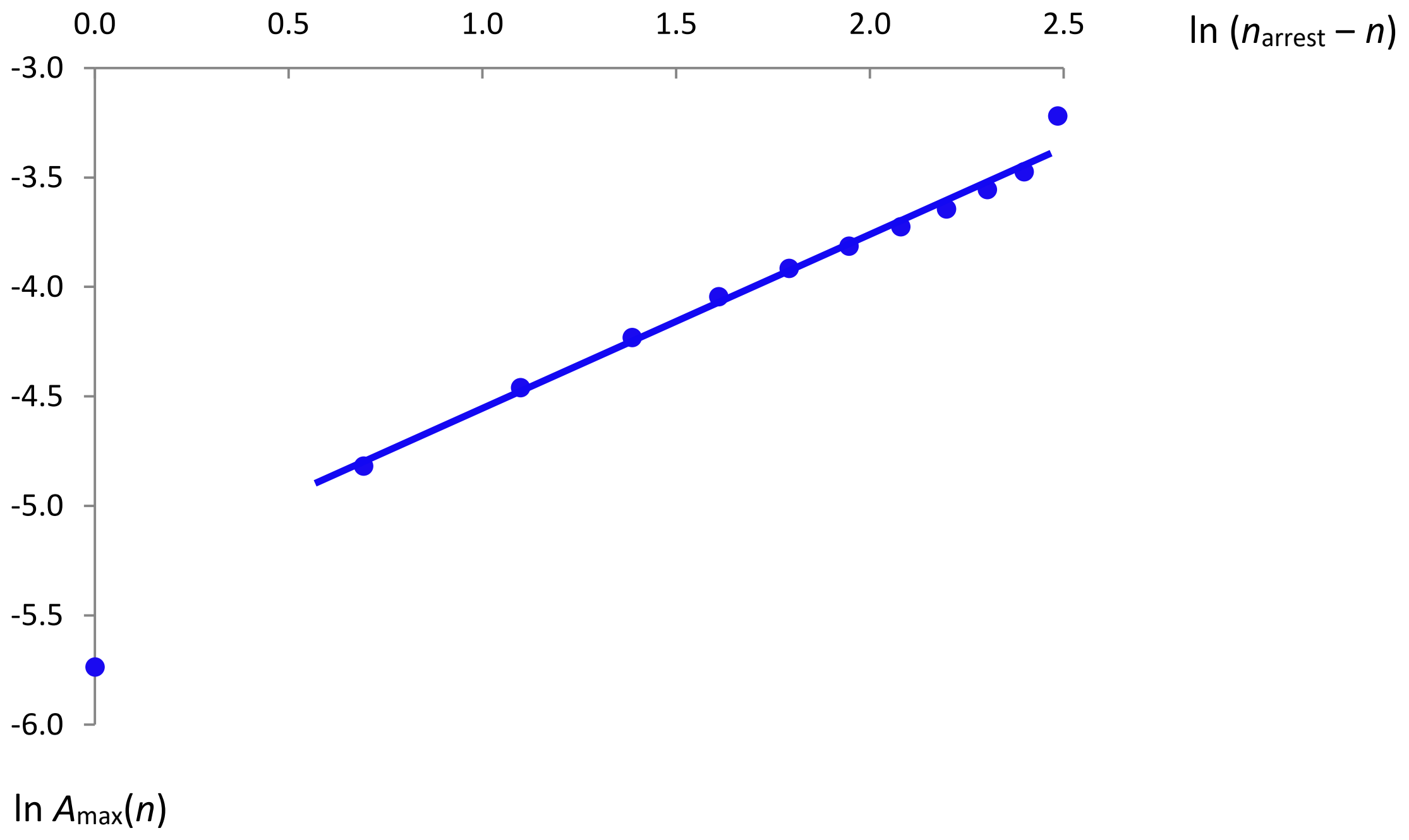

**Figure 13.** Travelling breather maximum amplitude $A_{max}(n)$ vs. $(n_{arrest} - n)$ in logarithmic coordinates. Nonlinear chain of damped linear oscillators with cubic coupling between the nearest neighbours. Numerical simulations with initial velocity amplitude $A = 0.04$ and viscous damping coefficient $\lambda = 0.00005$. Travelling breather arrest depth $n_{arrest} = 13$.

Finally, the hyper-exponential decay of the breather amplitude in the post-arrest regime is illustrated for $\delta = 3$ in Figure 14 (similarly to Figure 6).

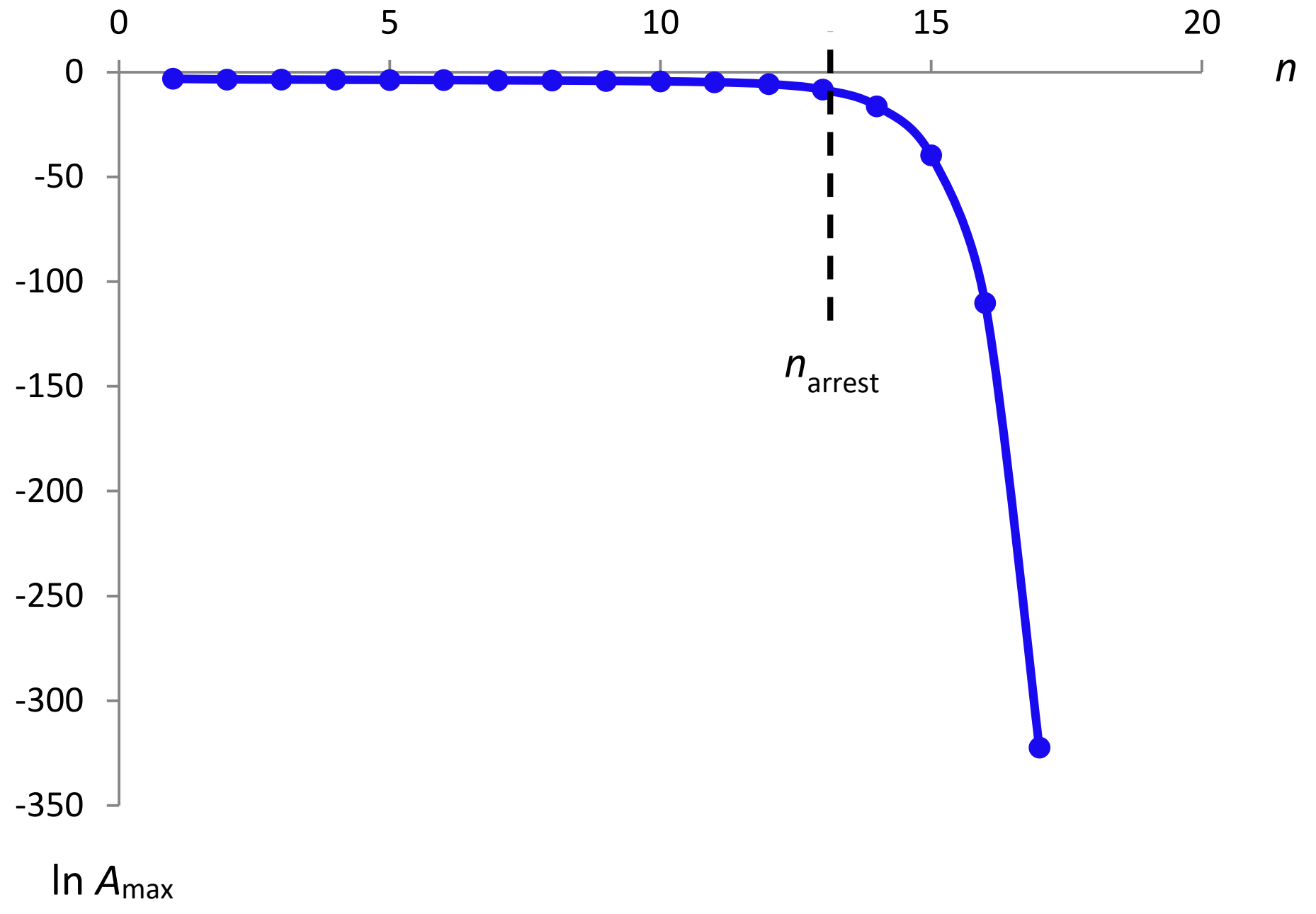

**Figure 14.** Logarithm of the maximum amplitude $A_{max}$ of the travelling breather vs. corresponding oscillator $n$. Nonlinear chain of damped linear oscillators with cubic coupling between the nearest neighbours. Numerical simulations with initial velocity $A = 0.04$ and damping coefficient $\lambda = 0.00005$. Travelling BA depth $n_{arrest} = 13$.

## 4. Conclusions

First of all, the results presented above relate the observed BA phenomenon to abrupt transition between two qualitatively different stages of the breather propagation. The first stage, or pattern, is characterized by relatively slow (power-law) decay of the breather amplitude. The simplified model of two oscillators allows surprisingly deep insight into this stage and yields reasonable predictions of the decay power-law and scaling of the penetration depth with the initial excitation amplitude and the damping, see estimations (35). Explicit formulas like (31) also yield correct order of magnitude for the penetration depth, although a numeric coincidence is less impressive. Perhaps, it is already too much to expect from this oversimplified model.

The next stage of propagation is characterized by extremely small breather amplitudes that decay hyper-exponentially with the penetration depth. Such additional penetration is inevitable due to the coupling inside the chain, but, needless to say, cannot be explained with the simplified 2DOF model. In any case, the amplitudes in this regime are by many orders of magnitude smaller than the initial excitation, and therefore this propagation stage hardly can be observed in any imaginable experiment.

The results allow one to conjecture that the BA phenomenon reveals itself due to the peculiar interaction of two generic factors – the damping and the essential coupling nonlinearity. Thus, one can expect to encounter similar pattern of the breather propagation in much broader class of possible models. One can also hope that the 2DOF basic model will help to understand main scaling relationships at the first penetration stage. In the same time, analytic approximation (26) can become insufficient, for instance, for symmetric coupling force $f(w) = f(-w)$. In this case, more refined analysis will be required.

Realistic experimental settings of the systems with Hertzian contacts, or with other strongly nonlinear coupling, usually involve some pre-compression, or some small linear component. It means that for such systems, low-amplitude dynamics becomes linear or quasilinear. Therefore, it might be impossible to observe experimentally the hyper-exponential decay of the breather amplitude in such systems. In the same time, for higher amplitudes, when the dynamics is still primarily governed by the nonlinear coupling, one can hope to observe the dependencies similar to (35).

Still, it is easy to imagine perfectly realistic model that delivers arguably the simplest possible and the clearest illustration of the BA phenomenon. It is the chain of damped linear oscillators of equal masses with clearance, when the excitation is transferred between the neighbours through impacts. It is easy to understand that the initial excitation will pass from site to site with decreasing maximum velocity, up to the point when the energy of the oscillator will be insufficient to cover the clearance and to pas the excitation to the neighbour. Then, the penetration depth in this case is rigorously finite, without the stage of hyper-exponential decay.

**Acknowledgements**

The Authors are grateful to Israel Science Foundation (Grant no. 1696/17) for the financial support of the present work.